\documentclass[journal=nalefd,manuscript=letter]{achemso}

\usepackage[version=3]{mhchem} 

\author{Jariyanee Prasongkit}
\author{Anton Grigoriev}
\author{Biswarup Pathak}
\affiliation[Uppsala University]
{Condensed Matter Theory Group, Department of Physics and Astronomy, Box 516, Uppsala University, SE-751 20 Uppsala, Sweden}
\author{Rajeev Ahuja}
\affiliation[Uppsala University]
{Condensed Matter Theory Group, Department of Physics and Astronomy, Box 516, Uppsala University, SE-751 20 Uppsala, Sweden}
\alsoaffiliation[Royal Institute of Technology]
{Applied Materials Physics, Department of Materials Science and Engineering, Royal Institute of Technology (KTH), SE-100 44 Stockholm, Sweden}
\author{Ralph H. Scheicher}
\email{ralph.scheicher@fysik.uu.se}
\affiliation[Uppsala University]
{Condensed Matter Theory Group, Department of Physics and Astronomy, Box 516, Uppsala University, SE-751 20 Uppsala, Sweden}

\title[Conductance of nucleotides in a graphene nanogap]
{Transverse conductance of DNA nucleotides in a graphene nanogap from first principles}

\begin{document}
\begin{abstract}
The fabrication of nanopores in atomically-thin graphene has recently been achieved and translocation of DNA has been demonstrated. Taken together with an earlier proposal to use graphene nanogaps for the purpose of DNA sequencing, this approach can resolve the technical problem of achieving single-base resolution in electronic nucleobase detection. We have theoretically evaluated the performance of a graphene nanogap setup for the purpose of whole-genome sequencing, by employing density functional theory and the non-equilibrium Green's function method to investigate the transverse conductance properties of nucleotides inside the gap. In particular, we determined the electrical tunneling current variation at finite bias due to changes in the nucleotides orientation and lateral position. Although the resulting tunneling current is found to fluctuate over several orders of magnitudes, a distinction between the four DNA bases appears possible, thus ranking the approach promising for rapid whole-genome sequencing applications.
\end{abstract}


The prospect of finding an improved method for whole-genome analysis
is driving significant research efforts to reach that goal. Over the
past decade, the traditionally used Sanger method has been
increasingly transformed into a highly parallelized and automated
process, enabling the rapid rise in decoded DNA sequences seen
today. Effectively, the \$10,000-genome has been reached through
this next-generation sequencing technology. However, for a truly
widespread deployment of DNA sequencing (e.g., in clinical trials
and eventually even for so-called personal medicine), cost and complexity of
the sequencing process will have to be reduced even further, in
order to arrive at a cost of \$1,000 or less per genome.

In an attempt to realize such third-generation sequencing
technology \cite{Schadt2010}, nanopores have been at the center of the research focus.
Initially, only the monitoring of the ionic current through biological pores was considered, and the merits of this approach continue to be actively investigated \cite{Derrington2010}. However, solid-state nanopores have become more and more attractive for the purpose of DNA sequencing \cite{Storm2005,Iqbal2007,Dekker2007,Wu2009,Taniguchi2009}, since they generally provide
better stability and can be more easily controlled \cite{Branton2008} than biological pores. Furthermore, instead of measuring the ionic current, it was suggested to outfit solid-state nanopores with embedded electrodes and instead monitor
the transverse tunneling current induced by them. This possibility
was at first only explored theoretically \cite{Zwolak2005,Lagerqvist2006,HaiyingHe2008,YuhuiHe2010}, because the technical challenges to outfit the nanopore with sufficiently thin embedded electrodes had prevented its actual fabrication until very recently \cite{Edel2010}.

About a year ago, a new suggestion was put forward \cite{Postma:2010} to use
graphene nanogaps in a double function as both separating membrane
and electrodes, solving the problem of alignment and making the
electrodes atomically thin \cite{Novoselov2004,CastroNeto2009} for optimal single-base resolution. Even
more recently, it was experimentally demonstrated for graphene nanopores \cite{Drndic2008,Zettl2009} that it is
possible to detect translocation events of DNA \cite{Dekker2010,Drndic2010,Golovchenko2010}. Furthermore, at least one density-functional-theory-based study explored the capabilities of a graphene nanopore setup for the purpose of distinguishing between nucleotides \cite{Prezhdo2010}. In our investigations, we have used state-of-the-art first-principles
methods to study the transport properties of nucleotides inside a
graphene nanogap, to assess whether or not this setup could be useful for
the purpose of DNA sequencing.

To this end, we investigated the tunneling transport
properties of the four nucleotides deoxyadenosine monophosphate
(dAMP), deoxythymidine monophosphate (dTMP), deoxyguanosine monophosphate (dGMP), and
deoxycytidine monophosphate (dCMP) when located between graphene electrodes with armchair edges chemically
passivated by hydrogen \cite{YuhuiHe:preprint2010}. The system is divided into three regions:
the left and right electrodes, and the central region, containing a
portion of the semi-infinite electrodes on either side of the
junction.

To construct the graphene-nucleotide-graphene system, we first optimized the
isolated nucleotide and graphene electrodes separately, and then proceeded to place
each nucleotide between the hydrogenated armchair edges of the semi-infinite graphene sheets. An
electrode-electrode spacing of 14.7 \AA\ (measured from H to H) is maintained throughout
the calculations, which allows each of the four nucleotides to be
accommodated within the gap in every possible orientation. The central region included graphene electrode edges on either side of the junction together with 8.65 \AA\ of each (left and right) graphene sheet in order to ensure that the perturbation effect from the nucleotides is
sufficiently screened. The left and right lead regions are constructed of periodically (${z\to\pm\infty}$) repeated graphene unit cells 15.62 \AA\ wide. Periodic boundary conditions along the electrode edges effectively create repeated images of the nucleotides separated by $\sim$10 \AA, which was found to be sufficiently large to avoid any unphysical interaction. The combined graphene-nucleotide-graphene system was then optimized again, allowing all
atoms in the central region to relax.

\begin{figure*}[ht]
\begin{center}
\includegraphics[width=16cm]{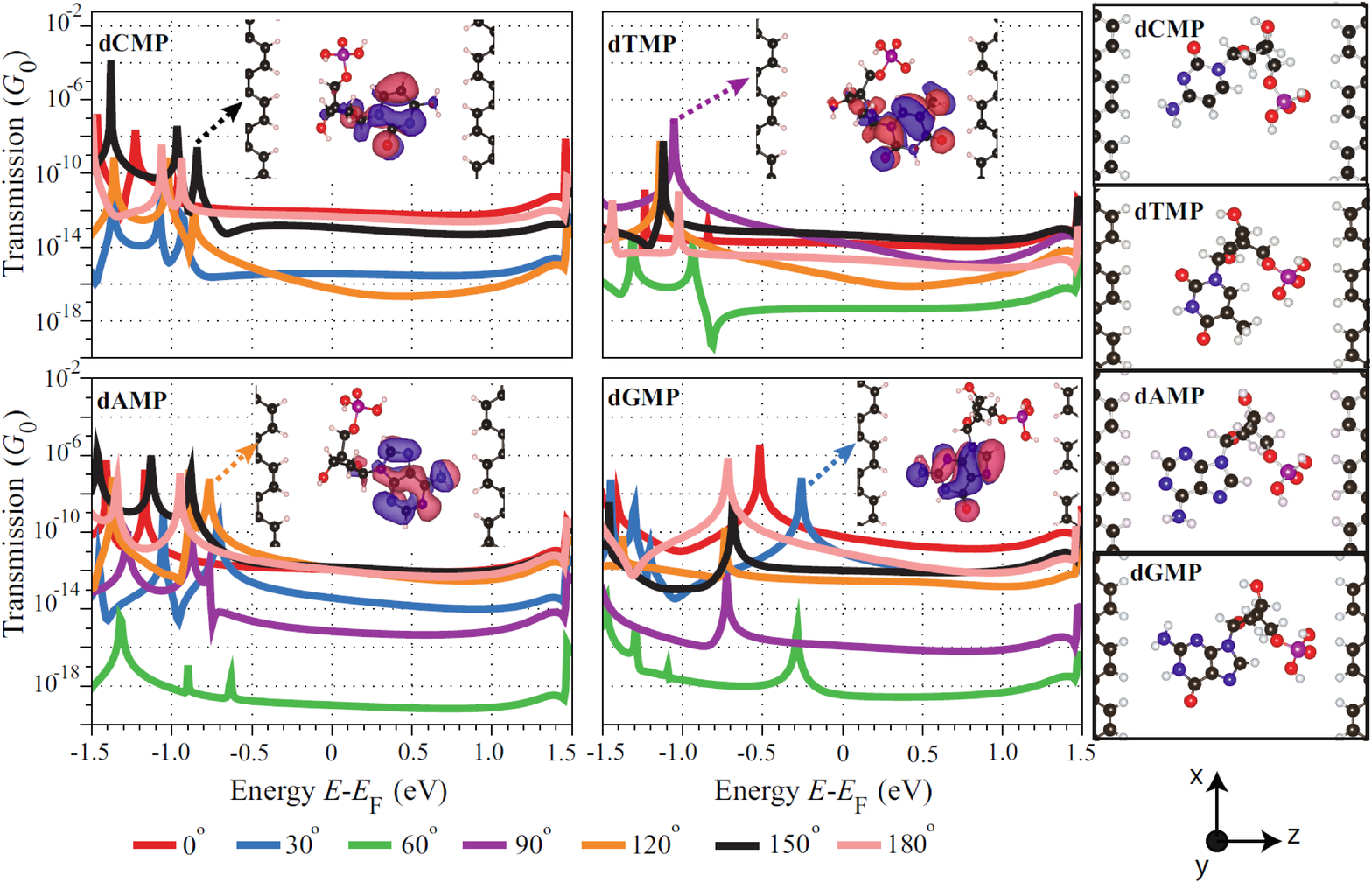}
\end{center}
\caption{The four central panels show the
zero-bias transmission function plotted on a semi-logarithmic scale
for the four nucleotides, dCMP, dTMP, dAMP, and dGMP. The respective
colors of the transmission curves indicate the angle by which the nucleotide has
been rotated in a counterclockwise direction around the $y$-axis
as per the legend at the bottom. The insets show isosurface plots of
the molecular orbitals responsible for those transmission peaks
marked by an arrow. The four vertically arranged panels to the right
display the nucleotide orientations corresponding to 0$^{\circ}$.}
\label{Transmission}
\end{figure*}

Each nucleotide was positioned so as to lie in the plane of the
graphene electrodes. We considered the effect of rotation and translation of the
DNA nucleotides on the transmission (see \ref{Transmission}). In our
investigation, the nucleotides are rotated around the $y$-axis from
0 to 180$^{\circ}$ in steps of 30$^{\circ}$ (Supporting Information), and translated along
the $z$-axis by $\pm$ 0.5 \AA\ for dGMP and dAMP and by $\pm$1 \AA\
for dTMP and dCMP.

All optimizations were performed by using density functional theory (DFT) as implemented in the SIESTA package \cite{Soler2002}. For the exchange and correlation functional, we employed GGA \cite{Lee:1988}. In the choice of the basis size, we follow Refs.\ \citenum{Soler2003} and \citenum{Soler2006}, adapting them for the Green's function type simulation: we found that a reduction from double- to single-$\zeta$ basis on C (and N) atoms does not affect the relaxed structure of the nucleotides placed between the graphene electrodes. Additional polarization functions were used in the case of P and H atoms. The atomic core electrons are modeled with Troullier-Martins norm-conserving pseudopotentials \cite{Troullier1991}. The real-space integration was performed using a 170 Ry cutoff, and due to the large cell size, only the $\Gamma$-point was considered for Brillouin zone sampling.

Transport calculations were carried out in the framework of the Landauer approach. We used the non-equilibrium Green's function (NEGF) technique based on DFT as implemented in the SMEAGOL package \cite{Rocha:2005,Rocha:2006}. The basis sets and the real-space integrations employed in the transport calculation are identical to those described above for the geometrical optimization part. The obtained transmission spectrum corresponds well with the test calculation performed with all double-$\zeta$ basis. The electric current $I$ was then obtained from integration of the transmission spectrum,
\begin{equation}
I(V_{b}) =
\dfrac{2e}{h}\int_{\mu_{R}}^{\mu_{L}}T(E,V_{b})[f(E-\mu_{L})-f(E-\mu_{R})]
dE, \label{equation1}
\end{equation}
where $T(E,V_b)$ is the transmission probability of electrons
incident at an energy $E$ from the left to the right electrode under
an applied bias voltage $V_b$, and $f(E-\mu_{L,R})$ is the
Fermi-Dirac distribution of electrons in the left (L) and right (R)
electrode with the respective chemical potential $\mu_{L} =
E_{\mathrm{F}}+V_b/2$ and $\mu_{R} = E_{\mathrm{F}}-V_b/2$ shifted
respectively up or down relative to the Fermi energy
$E_{\mathrm{F}}$. Further details of the method are described in
Ref.\ \citenum{Datta:1995}.

During the translocation process of a DNA strand through the
graphene nanogap, many different orientations of the nucleotides are
possible with respect to the graphene electrodes. It is therefore
crucial to consider how the transmission function depends on the
orientation of a given nucleotide. As seen in \ref{Transmission}, for all nucleotides, the Fermi level is aligned
near the HOMO. The isosurface plots of the molecular orbital
corresponding to the first transmission peak below the Fermi energy
are found to be associated with the HOMO of isolated bases,
localized on the pyrimidine and imidazole rings, as shown in the
respective insets. Consequently, the position of the resonance
peak and transmission values are significantly correlated with the
base orientation. When a nucleotide is rotated, the peak position
shifts upward or downward relative to the Fermi energy, and the
transmission changes its magnitude.
This is a result of the nucleotide-to-electrode coupling change. The transmission drops exponentially when
the nucleotide-graphene coupling is weakened due an increasing distance
of a nucleotide from the graphene edge. Zero transmission occurs in the
case when a nucleotide is so far removed from the graphene leads that virtually no
overlap between the states localized on the nucleotides and the
states on the graphene lead exists, as it can be seen from the absence of transmission curves in \ref{Transmission} for certain
orientations of dCMP and dTMP.

Let us first discuss how the zero-bias transmission function $T(E,
V$=$0)$ is influenced by the different base types. There are two
groups of nucleotides: one containing purine bases (A and G), and
the other containing pyrimidine bases (C and T). The main
distinctive physical property between these two groups is
the base size: purines are larger than pyrimidines. This leads to smaller separation and stronger
coupling of dGMP and dAMP with the electrodes. For all orientations
considered here, $T(-1~\mathrm{eV} \leq E \leq 1~\mathrm{eV},~
V=0)$ of dGMP and dAMP ranges from 10$^{-20}$ to 10$^{-6}$ $G_0$,
while the corresponding transmission function of dCMP and dTMP can
range from 0 to 10$^{-8}$ $G_0$. The case of zero transmission for
dCMP and dTMP is obtained at certain orientations when no coupling
exists between the nucleotides and the leads.

Different molecule-electrode separations, and especially effective localization of the HOMO in the middle or at one side of the inter-electrode gap have an immediate effect on the width of the HOMO resonance. The peak widths of the HOMO resonance exhibit a variation due to different base orientations, where the peak widths
of dGMP and dAMP are $\sim$ 0.10--0.20 eV, while those of dTMP and
dCMP are $\sim$ 0.05--0.10 eV. This narrowing of the HOMO peak width
is a result of weaker coupling between bases and leads for dTMP and
dCMP as compared to dGMP and dAMP, caused by narrowing of the transmission cone and by increased localization (and lifetime) of the electron in the HOMO coupled to the electrodes. Together with the molecule-electrode separation, the nucleotide chemistry and corresponding HOMO symmetry and rotation plays an important role in this coupling and localization of the state.

Comparing the results for dGMP and dAMP, we note that the Fermi
level is aligned very closely to the HOMO peak of dGMP ($E \leq
-0.5$ eV), while that of dAMP appears at $E \geq -0.5$ eV. This
circumstance gives rise to a significant difference in conductivity
between dGMP and dAMP, which enables electrical distinction between
them at low bias voltages.

Upon comparison of the transmission functions of dCMP and dTMP, we
see that there is not much difference in the magnitude and position
of the resonance peaks (at $E \leq -0.6$ eV) for a wide range of
orientations, and thus it would seem at first that these two
nucleotides cannot be easily distinguished. However, as we shall see below, it is in fact possible to achieve an unambiguous distinction once the effects due to finite bias are considered.

\begin{figure}[ht]
\begin{center}
\includegraphics[width=8cm]{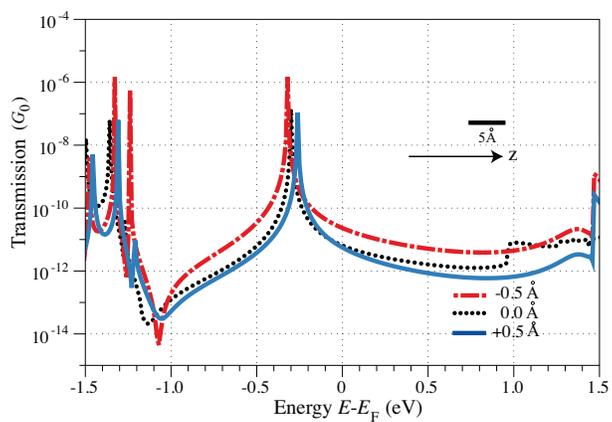}
\end{center}
\caption{The change in transmission function of dGMP
(with an orientation corresponding to a 30$^{\circ}$ rotation, as
defined in \ref{Transmission}) due to translation in steps of
$\pm$0.5\AA\ along the $z$-axis.} \label{translation}
\end{figure}

To understand how the transmission is affected not only by rotation
but also by lateral translation of the nucleotides, we tested effects of a
shift in position within the $x$-$z$-plane. \ref{translation}
presents the resulting change in the transmission function with
translations of dGMP in steps of $\pm$0.5 \AA\ along the $z$-axis. Note, that the rotation of the nucleotide alone is coupled to the effective localized HOMO translation between the electrodes. We find that the width of the HOMO resonance peak increases when the
base part of the nucleotide is moved closer to the lead due to a broadened transmission cone and decreased lifetime of the coupled state. The
broadening of the transmission peak width results from a stronger
coupling between base and graphene electrode.

A shift of the HOMO peak position is observed both for rotation and translation of the nucleotide relative to the graphene edges.
 This is due to Pauli repulsion of the states on the nucleotide and the electrode edge, however charging of the phosphate moiety's (known to act as an electron acceptor) may play a role as well.
 An increased accumulation of electrons on the phosphate-group leads to an overall charging of the nucleotide, slightly shifting the molecular energy levels towards lower energies. For all nucleobases, the behavior of the shift and the peak width was found to be qualitatively the same, however, for the smaller pyrimidine bases dCMP and dTMP, the resonance peak can shift by up to 0.2 eV.

From the zero-bias transmission functions, we conclude that the difference of chemical and physical structures between the purine bases (dAMP, dGMP) and pyrimidine
bases (dCMP, dTMP) affect the coupling strength of the DNA bases with the graphene electrodes, thus leading to the possibility to distinguish the two groups of DNA nucleotides under applied bias.

\begin{figure}[ht]
\begin{center}
\includegraphics[width=8cm]{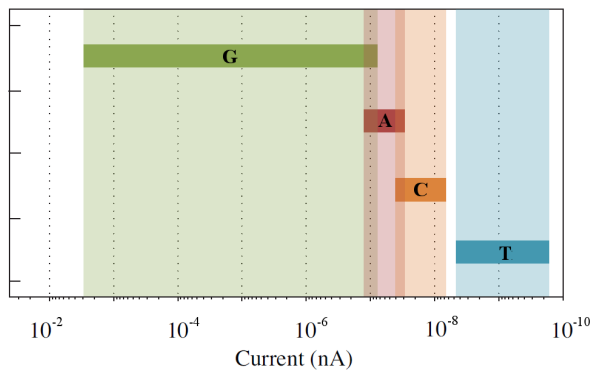}
\end{center}
\caption{Current variation due to nucleotide rotation about the
$y$-axis and translation along the $z$-axis at a bias of 1 V.}
\label{ibar}
\end{figure}

\ref{ibar} shows the range of possible current values for a
bias voltage of 1 V when the nucleotides are rotated and
translated again in the manner discussed above. The large fluctuations in
the current are caused by the large variation of nucleotide-graphene
coupling strength. It is important to emphasize that not too much
attention should be paid to the absolute value of the current, which
is provided in \ref{ibar}; rather, the ratio of the currents
may be regarded as more relevant. The absolute current value is
largely influenced by our simulation settings, where we use
truncated nucleotides instead of extended DNA chains. In fact, we
found from a test that when further parts of the sugar-phosphate
backbone are included in the simulation, the current does in fact
increase. Also, the current could potentially be increased by more
than one order of magnitude trough the creation of a low
concentration of impurities in the graphene electrodes; such a
conductivity-raising effect was recently demonstrated from both
experiment and theory \cite{Jafri2010,Carva2010}.

The magnitude of the currents is seen to be ordered in the following
hierarchy: $I_{\mathrm{dGMP}} > I_{\mathrm{dAMP}} > I_{\mathrm{dCMP}} > I_{\mathrm{dTMP}}$. Thus, dGMP can be
distinguished from the other nucleotides due to its strong broad
current signal which results from the Fermi energy of the graphene electrodes being
close to the wide HOMO peak of dGMP. The other three nucleotides (dAMP,
dCMP, and dTMP), which possess HOMO peaks further away from the graphene electrodes
Fermi energy, exhibit different characteristic current
magnitudes, showing rather little overlap with each other.
In our analysis, we neglected very low current values below $I <
10^{-11}$ nA, which are expected to disappear into the electrical
background noise in experimental measurements.

From the viewpoint of DNA sequencing applications, it is on the one hand
encouraging to see that dAMP and dCMP exhibit a relatively narrow
current range which should make them easier to identify. On the
other hand, dGMP and dTMP have relatively broad current ranges,
covering several orders of magnitude. However, for dGMP, the current
is always at the higher end, while for dTMP, the current is always
at the lower end of the scale. Thus, based on our results, it should
be in principle possible to distinguish between all for nucleotides
in the graphene nanogaps setup.

\acknowledgement

We thank Henk Postma for very helpful discussions throughout the course of this project.
Financial support is gratefully acknowledged from the Royal Thai Government, Carl Tryggers Foundation, the
Uppsala University UniMolecular Electronics Center (U$^3$MEC),
Wenner--Gren Foundations, and the Swedish Research Council (VR, grant no.\ 621-2009-3628). The
Swedish National Infrastructure for Computing (SNIC) and the Uppsala
Multidisciplinary Center for Advanced Computational Science (UPPMAX)
provided computing time for this project.

\bibliography{achemso}


\newpage

\suppinfo
Figures showing the different orientations of the four nucleotides in the graphene nanogap.

\end{document}